\documentclass{www13-companion-accepted}
\usepackage{graphicx}
\begin{document}
%

\title{{\ttlit Serefind:} a Social Networking Website for Classifieds\titlenote{http://serefind.com}}

%
%
%
%
%

\numberofauthors{1} 
%
\author{
%
Pramod Verma\\
\affaddr{Serefind}\\
\affaddr{13 S Regester St, Baltimore, MD, 21231, USA}\\
\email{founder@serefind.com}
}

\maketitle
\begin{abstract}
This paper presents the design and implementation of a social networking website for classifieds, called \textbf{Serefind}. We designed search interfaces with focus on security, privacy, usability, design, ranking, and communications. 
We deployed this site at the Johns Hopkins University, 
and the results show it can be used as a self-sustaining classifieds site for public or private communities.
\end{abstract}

\section{Introduction}
Let us consider the typical method of interaction between buyers and sellers in classified settings. 
When people want to sell something, they advertise by way of a physical or virtual announcement. 
For example, an individual puts up a flyer in a neighborhood to attract local buyers. 
People who are interested in the classified listing contact the owner either by phone or email. The two parties communicate until the deal is consummated or falls through. In this paper, we address problems related to security, privacy, design in available classifieds websites, and then propose design and implementation of Serefind: A social networking for classifieds.


\section{Background}
The growing prevalence of Internet access has enabled classified communications to emerge online. 
There are hundreds of web-based services for this kind of classifieds communication. For the sake of simplicity, we will only review top classifieds websites according to alexa-web rank.
One popular example is Craigslist\footnote{http://craigslist.org}, which according to alexa-web rank, 
is ranked 45th in the world. 
This website contains classified data on a large scale. 
It has various problems related to security, 
privacy, design, and usability. There exist hundreds of cases showing 
that use of this website has lead to crimes such as kidnapping, threats, and prostitution. 

Another example is EBay Classifieds\footnote{http://ebayclassifieds.com} which also facilitates 
classified transactions. Some sites such as Amazon are e-commerce websites where people can buy items, 
but these have a different structure than classified websites.
Some social networking websites such as Facebook provide a framework for third-party 
classified applications to leverage an existing social graph. Facebook Marketplace\footnote{http://apps.facebook.com/marketplace/} powered by Oodle, a third-party company, is an example of this kind of application. 
Third party association creates various privacy related issues.
\section{Design Goals}
Our design goal aimed to create a social classifieds site that satisfies the following constraints:
\subsubsection*{Security and Privacy}
The identity of anyone who uses the website should be verified to reduce crime. 
For example, on Craigslist, interactions can occur without thorough 
verification on either side. If an issue arises, since there is no verification, 
it is difficult to trace the individuals involved. Furthermore, 
all websites should strive to meet the web based security standards.
\subsubsection*{Social-Graph}
Today we live in the social networking era where we conduct social experiences online via computers, 
hand-held devices, and gaming consoles. Social networking is changing our way of communication, 
allowing us to easily share information with our friends, family and peers. 
Online marketplaces can be visualized using various social graph models\cite{Kumar:2007:SNA:1336000.1336197}. 
Social network and commerce networks are interconnected to each other\cite{Guo:2011:RSN:1993574.1993598}.
\begin{figure}
  \begin{center}
   \includegraphics[width=\columnwidth]{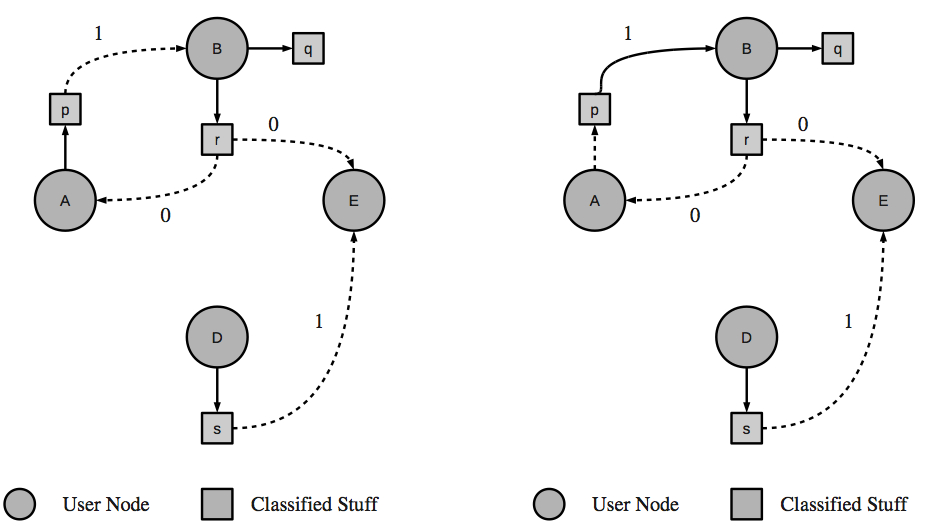}
  \label{fig:socialgraph}
    \caption{In figure, user $A$ owns a classified stuff $P$ (solid arc between $A$ and $P$) and user $B$ is communicating for that item (dashed arc between $P$ and $B$).
Arc between $B$ and classified node $P$ becomes solid and Arc between $A$ and $P$ becomes dashed, because it has been sold from user $A$ to $B$.
  }
  \end{center}
\end{figure}
To make it simple, we present a social graph that can be visualized using a graph with nodes and edges. 
Node represents be owner (circle) and classifieds stuff (square).
 A solid edge represents ownership of a listing, and a dashed edge represents interest in a listing from a peer. 
 Total message count for communication regarding a listing is represented with a number above the edge.
One of the design goals is to follow the above social graph model (See figure~1). 
\subsubsection*{Design, HCI and User Interfaces}
Sites should use suitable Human-Computer Interaction techniques to enrich the user experience. 
Users should be able to provide information in a user-friendly manner. Classified data should be browsed 
with usable search interfaces. Communication between users should be easy and efficient. As for advertisements, minimal use of banner and text ads would enhance the user experience. 
In addition, the site should be aesthetically pleasing, on par with or more so than existing classifieds sites, while maintaining an aura of safety and minimalism.



\subsubsection*{Ranking Algorithm}
The site should use a sophisticated ranking algorithm to display appropriate and relevant results from a large database.
For instance, we can incorporate serefind social graph elements such as interactions and communications to provide an improved ranking algorithm.

In the next sections, we will present the design and implementation of site by describing user-interfaces, ranking algorithms, and the design of communications.
\section{Serefind System}

\begin{figure*}
  \begin{center}
    \includegraphics[height=29.5mm]{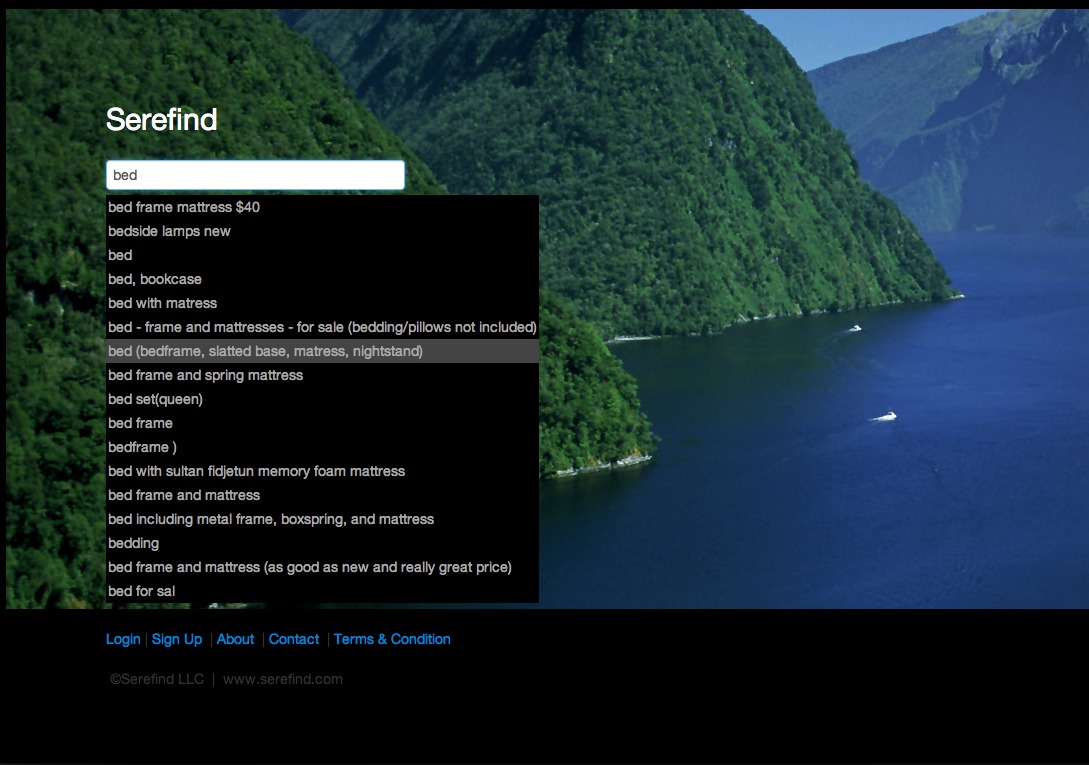}
    \includegraphics[height=29.5mm]{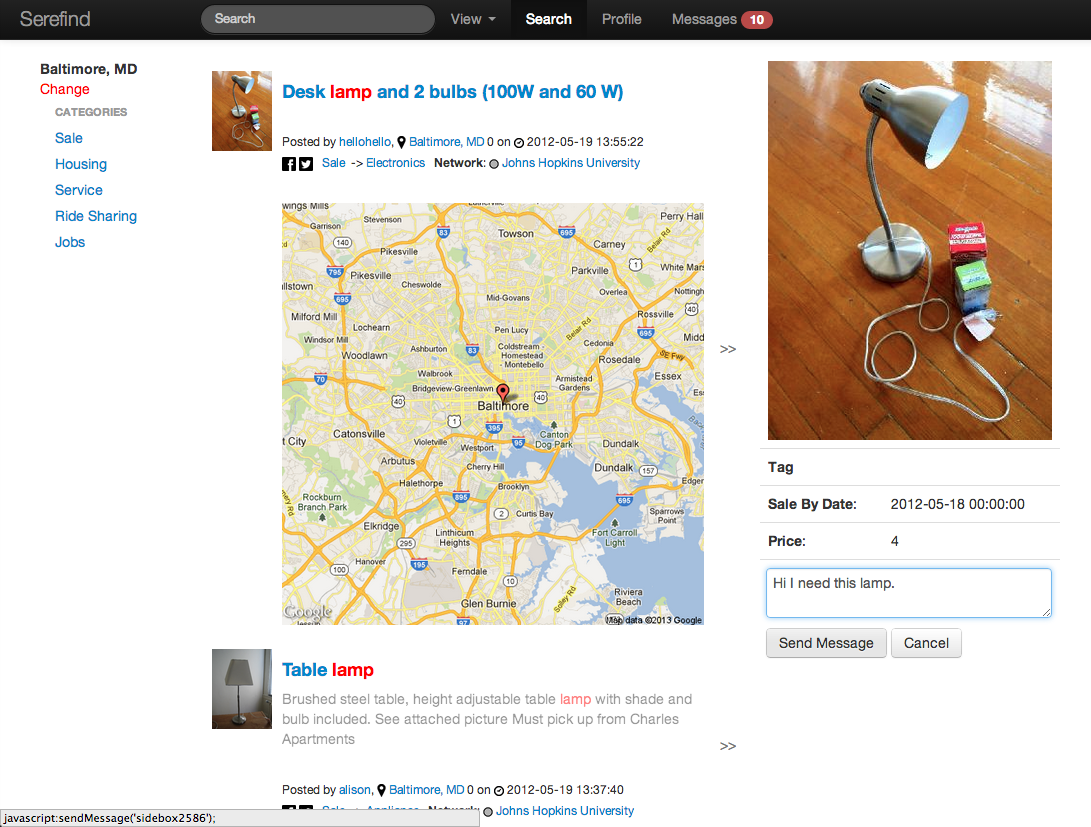}
    \includegraphics[height=29.5mm]{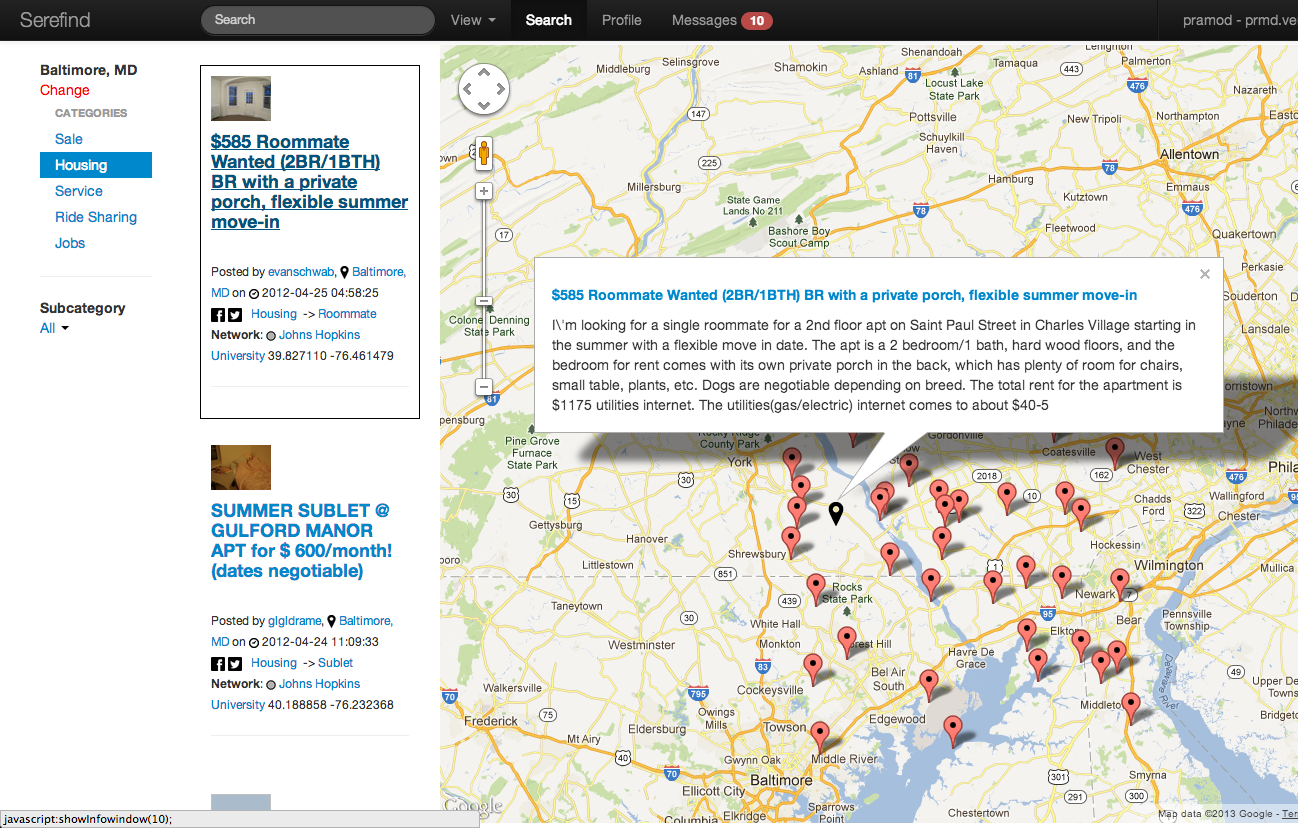}
    \includegraphics[height=29.5mm]{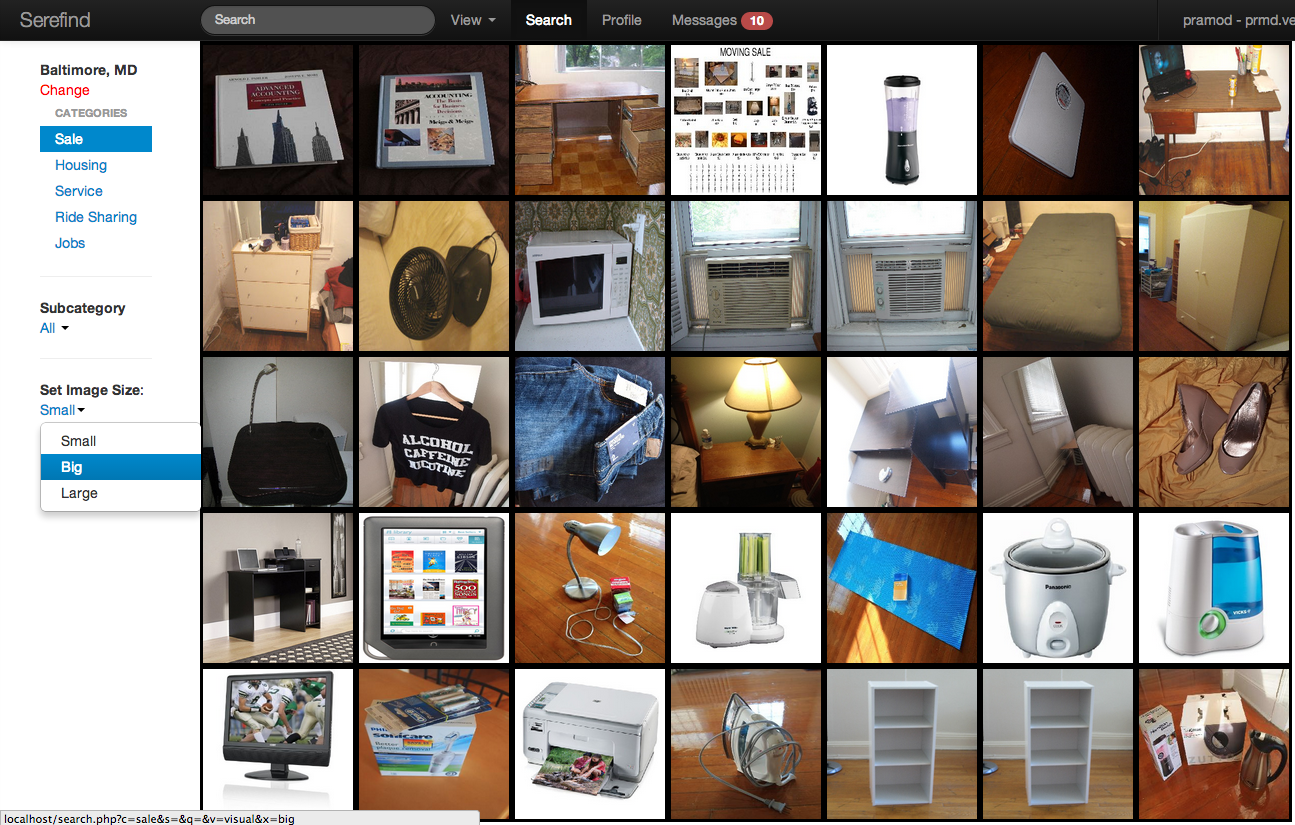}
  \caption{(Images from left to right) 
  (1) Landing page of Serefind with instant search menu, 
  (2) Home page with navigation bar (top), search interface, categories, location settings, instant view with message UI and news feed, 
  (3) Map view (4) Visual search (zoom for details or visit http://serefind.com/demo).}
  \label{fig:serefind}
  \end{center}
\end{figure*}

Figure~\ref{fig:serefind} shows the Serefind user-interfaces. Serefind has three main user-interfaces 1) 
Search interface 2) Profile, and 3) Communication. In addition we have authentication, settings and 
help interfaces. Serefind users have to register with a well-defined email address affiliated with a network. 
For example the user with \textbf{pramod@cs.jhu.edu} will join \textbf{The Johns Hopkins University} network. 
This verification ensures security. The navigation bar at top of the site is used to access these 
sections. Users can set network, password and email settings.

After registration and login, the user visits the \textbf{Search Page} (default home page) that displays recently 
added classified listings in the form of a \textbf{Newsfeed}. This is important because it shows users 
that the site is being used. By default, 
the Newsfeed is displayed based on personal preferences, such as location, network, and categories. 
We also provide a link to customize preferences.
\subsection{Profile}
In a social network, a user's \textbf{Profile} is a place where the user adds and manages posts (classifieds listings in our case). 
Profile UI and its elements can be used to signal other user nodes and connections\cite{Lampe:2007:FFP:1240624.1240695}.
We had the following requirements for the Profile metaphor.

\subsubsection*{Heterogeneous Data Sharing}
 To support all kind of heterogeneous classifieds data, we have developed 
a template based framework, where all the specification for categories can be written in the form of xml 
files. This approach assists to create new categories and their specification 
directly from users with administrative control (approval), without modifying the source code. 
Template also mentions specifications for the search. For example following code specifies \textbf{Event} category in the \textbf{event.xml} 
. Database stores information according to the XML definition. 
UI is displayed based on XML file. 
\begin{verbatim}
<schema id="O198" category="event" creator="admin">
	<field input-type="textbox"  data-type="text" 
	visibility-in-search-filter="true">Title</field>
	<field data-type="date-time">Date and Time</field>
	<field>...</field>
</schema>
\end{verbatim}		
This is a unique approach compare to the existing websites such as Facebook, Twitter, etc., 
which only allow a user to post specific types of information. 
In theory, Serefind can be used to post any type of information beyond classifieds even tweet and 
Facebook like post. We just have to put related XML file with specifications in the system, and 
Serefind search and post interfaces will be changed automatically. 
In addition, We also implemented an interface in which users can request new categories and their respective fields. After approval from administrator, new categories can be part of Serefind without source code modification. 
\begin{verbatim}
	<requestField category="event" data-type="currency" 
	creator="user001">Cover Charge</requestField>
\end{verbatim}
\subsubsection*{Security and privacy}
By default, at the Profile page, we do not display full name, location, photo, or any other information about a user. 
We simply display a username and set of listings. 
We do not show any messages regarding these listings. 
If the user is not signed in, most details such as description, user name are hidden by default. In addition, 
users can explore listings belonging to their own network. However public listings are visible to all the networks.
\subsubsection*{Post listings}
By default post interface shows HTML input elements for category, subcategory, tags, title, and description. The user can enter data in profile according to which classified category the data falls under. The Profile is designed using intelligent interactive methods to make data entry and repetitive data entry easier and more usable. 
After data is entered, addition of the entry is confirmed and the user is immediately able to add another listing. Users can add more input-fields according to their need.

\subsubsection*{Manage Listings}
The user can manage listings by using the edit, delete, or hide links provided for each listing. We also provide an undo button if the user deletes or hides a listing by mistake.
\subsubsection*{Navigation}
For a user to view the profile of another user on the lightbox and listing page, we provide a link to the owner's profile. 
Another way to access a specific user profile is to visit http://serefind.com/directory/profile/[username].
\subsubsection*{Other Interfaces}
For a user to determine if their listings are being viewed, we display the number of views for each 
listing when the user visits profile. This allows the users to see how many people are viewing their listings 
and perhaps serves as impetus to make changes if the listing is not being viewed or the 
listing is being viewed and no responses are being generated.
\subsection{Search Interfaces}

Search interfaces play an important role for any data-oriented website or system. Researchers have attempted to improve conventional interfaces \cite{Xu:2009:NVS:1498759.1498821}.
We set the following requirements for Serefind to ensure optimized design and usability.

\subsubsection*{Data Visualization}
Serefind is a data driven site. We provide few classified categories such as 
books, furniture, apparels electronics, housing, sales, service, and jobs. 
Without proper visualization and interface, users cannot interpret data and interact regarding data. 
We choose the following user interfaces to visualize classified data:
\begin{enumerate}
\item  List View: Each listing contains basic information such as icon picture, title, 
user name of owner, description, network, category, subcategory, and date posted. More details is displayed in the left side of listings on mouse over on expand icon.
\item  Thumbnails View: This is a visual search in which we only display pictures and title.
\item  Map View: We used the Google Maps API to visualize location based classifieds data. 
This feature is useful when a user is searching for data in a particular location and is also useful in increasing the relevancy of 
data by showing listings which are nearby to the user.
\item Tabular View: In this view, results are displayed in a tabular format so that user can easily view and sort.
\item Item Pages: These pages contain the full set of details regarding a listing. A lightbox appears on the right side of the page when a user selects a listing by clicking on the title or picture. This lightbox method allows the users to preview the listing and decide whether he is interested in
pursuing it further. It also provides a link to the listing page which contains complete details of that classified listing.
\item Lightbox: This is a simple overlay that appears when a user selects a listing. This helps to explore details 
without redirecting to the item-page.
\end{enumerate}

\subsubsection*{Page Scrolling}
We use AJAX based infinite-scrolling method to see more results. During the scrolling, top and side navigation bars remain fixed.

\subsubsection*{Sorting and Filters}
We display category menus and search filters on the left side of the site. Tabular view also displays column-links for sorting the results.

\subsubsection*{Navigation}
How can a user reach appropriate results? We designed the interface in such a way that users 
can access information in three steps. In the first step, a query is invoked. The second stage is the
browsing step, in which the search results are represented in a set of relevant listings displaying name, title, price, and description, and 
the user can browse these listings and select those which he is interested in. 
The third step involves the display of a lightbox when users select a listing. 
At this point, the user can communicate regarding a listing or go to the listing page to view more information.

\subsubsection*{Communication} For users to communicate with owners during a search, 
We placed a message box at the bottom of the lightbox.
\subsubsection*{Other Usability Factors} How we ensure usability? We considered many usability-related problems 
for search interaction. For instance, to minimize user time and effort, we introduced an instant search on each page. 
Another simple usability problem is how the user can view multiple listings in an efficient manner. 
We use a lightbox or instant overlay at the right-hand side of the page so the user can view N listings with N clicks.
\subsubsection*{Ranking Algorithm}
Our social classifieds system can be represented with a graph where owners and their 
classifieds listings are represented respectively as nodes and their edge relationships 
are described as a foreign-key in their respective relational database schemas.
 
A classifieds  $C$ can be represented as $C(O, c_{1},...c_{k})$ where $O$ is the owner and $c_{i}$ is the 
attribute such as title, description, price, username, etc. 
Search query can be represented as $Q(U, q_{1}, q_{2}, q_{3}...q_{n})$ where $U$ is the user who 
invokes the query and $q_{i}$ is the search term(s). The similarity score between a query term $Q$ and 
classified nodes $C$ can be represented by $Sim(Q,C) = \sum S(q_{i},C)$, which can be calculated by various 
Information-Retrieval (IR) based methods such as TF- IDF and RF \cite{QwweeRank}. 
To provide better search experience, we used Location Based Service (LBS) and Query Expansion (QE)\cite{Carpineto:2012:SAQ:2071389.2071390} methods to optimize the search results.
\subsection{Design of Communication}
Communication is the basic step for any classifieds website or social networking website. 
On popular websites such as Facebook, Gmail, there exists an \textbf{Inbox} metaphor for communication. 
Serefind Message section contains interfaces for inbox, sent, and deleted messages.
Our site was designed for classifieds, so the design of communication has the following 
constraints and requirements:
\subsubsection*{Initiating a Communication}
How does a user initiate communication? When users are searching, they can directly send a message to 
the owner using the send message UI at lightbox which appears when a user selects a listing. 
User can also send message on Item-Page. This creates a new message associated with a particular listing, thus automatically 
providing a context for each message exchanged.
\subsubsection*{Security and Privacy}
How we ensure privacy and security? Users can view and delete their own messages or comments. In addition, 
usernames are also in the form of a self-assigned moniker. 
Because each user has to register with a valid organizational email, 
we can track any issue or crime originating with a communication on the site.
\subsubsection*{Context of Communication} 
What will the subject line be? Traditionally, in an inbox-based system, users are allowed to message any other user without any context or subject line. We present listing title as a subject or context for the communication.
\subsubsection*{Notification}
How can users access new messages? 
Serefind generates email notifications with a link to the inbox. 
Here we faced and experimented two different design options. 
Should we send an actual message within the email notification or 
we should provide a small message to login to the site to read a message? 
In the first approach, instead of reading a full message with a link for replying message, 
a few users sent the message to support@serefind.com because after reading messages 
they were confused about the identity of the sender. The second approach seemed solid but slightly 
less convenient, because users have to login to the site to read complete messages.
Within the site, we provide a notification on the top bar indicating number of new messages.
\subsubsection*{Special-Cases}
What happens when an owner deletes a listing? In this situation, users who previously 
communicated regarding a listing can view all past communications. However, users cannot send 
further messages to the owner of the deleted listing.
\section{Conclusion}
We implemented Serefind using open-source technologies such as PHP and MySQL. To provide better 
user experience, we are using AJAX to lessen page refresh time at various places in the site. 
The site was launched at Johns Hopkins University.
Currently site has approximately 10,000 users who belong to \textbf{The Johns Hopkins University} and 
nearby schools networks. At the time of the experiment, the website's alexa world web rank was 17,802,775 on 
a single network. 

We believe our work will be a better model of a classified system by focusing on design, 
usability, communication, search, and security. 
Serefind interface will create an optimized user experience for those interested in an 
alternative to the traditional classified system. 
A demo is available at \textbf{http://serefind.com/demo}
\bibliographystyle{abbrv}
\bibliography{sigproc}  
\end{document}